\documentclass[conference]{IEEEtran}
\IEEEoverridecommandlockouts
\usepackage{cite}
\usepackage{amsmath,amssymb,amsfonts}
\usepackage{graphicx}
\usepackage{algorithm}
\usepackage{algpseudocode}
\usepackage{textcomp}
\usepackage{url}
\usepackage{multirow}
\usepackage{tcolorbox}
\usepackage{booktabs}
\usepackage{tcolorbox}
\tcbuselibrary{skins}
\usepackage{multirow}
\usepackage{bbding}
\usepackage{utfsym}
\usepackage{ulem}
\usepackage{balance}
\usepackage{wasysym}
\usepackage{hyperref}
\usepackage{multirow}
\usepackage{makecell}
\usepackage{pifont}

\usepackage[utf8]{inputenc}
\usepackage[T1]{fontenc}

\def\eg{\textit{e.g.,} }
\def\ie{\textit{i.e.,} }

\newtcolorbox{AcademicBox}[1][]{academicbox=#1}
\definecolor{SoftBlue}{RGB}{135, 206, 250}  
\definecolor{SoftOrange}{RGB}{255, 224, 178} 
\definecolor{SoftGreen}{RGB}{144, 238, 144}  
\definecolor{CorrectGreen}{RGB}{76, 175, 80} 
\definecolor{ErrorRed}{RGB}{211, 47, 47} 

\def\BibTeX{{\rm B\kern-.05em{\sc i\kern-.025em b}\kern-.08em
    T\kern-.1667em\lower.7ex\hbox{E}\kern-.125emX}}

\tcbset{
  my box/.style={
    enhanced,
    colframe=#1!80,
    colback=#1!10,
    attach boxed title to top left={xshift=0.2cm, yshift=-0.2cm},
    boxed title style={
      colback=#1!80,
      outer arc=0pt,
      arc=0pt,
      top=0pt,
      bottom=0pt,
    },
  },
}

\begin{document}

\def\eg{\textit{e.g.,} }
\def\ie{\textit{i.e.,} }

\newcommand{\dm}[1]{{\color{blue}{\textbf{#1}}}}

\title{A Conceptual Framework for Requirements Engineering of Pretrained-Model-Enabled Systems}


 \author{
 \IEEEauthorblockN{Dongming Jin$^{1,2}$, Zhi Jin*$^{1,2}$, Linyu Li$^{1,2}$, Xiaohong Chen$^{3}$}
    \IEEEauthorblockA{$^1$ Key Laboratory of High Confidence Software Technologies (Peking University), Ministry of Education, China}
    \IEEEauthorblockA{$^2$ School of Computer Science, Peking University, China}
    \IEEEauthorblockA{$^3$ East China Normal University, China}
    \IEEEauthorblockA{\texttt{correspondence to: zhijin@pku.edu.cn}}
}


\maketitle

\begin{abstract}

Recent advances in large pretrained models have led to their widespread integration as core components in modern software systems. The trend is expected to continue in the foreseeable future. Unlike traditional software systems governed by deterministic logic, systems powered by pretrained models exhibit distinctive and emergent characteristics, such as ambiguous capability boundaries, context-dependent behavior, and continuous evolution. These properties fundamentally challenge long-standing assumptions in requirements engineering, including functional decomposability and behavioral predictability. This paper investigates this problem and advocates for a rethinking of existing requirements engineering methodologies. We propose a conceptual framework tailored to requirements engineering of pretrained-model-enabled software systems and outline several promising research directions within this framework. This vision helps provide a guide for researchers and practitioners to tackle the emerging challenges in requirements engineering of pretrained-model-enabled systems.

\end{abstract}





%

\section{Introduction}

With the recent breakthroughs in large pretrained models (\eg large language models and multi-modal models), these models are no longer mere auxiliary tools but are integrated as core components in modern software systems~\cite{weber2024large}. Within such systems, pretrained models typically serve as reasoning engines, knowledge bases, or intelligent agents~\cite{luo2025large}~\cite{jin2024mare}. We refer to such software systems as pretrained-model-enabled systems in this paper. These systems have already shown significant potential to impact various application domains, \eg social simulation~\cite{park2022social}, research assistant~\cite{ziems2024can}, and software engineering~\cite{dong2024self}. This trend is expected to continue in the foreseeable future.

Unlike traditional software systems governed by deterministic logic and rule-based behaviors~\cite{indykov2025architectural}, pretrained-model-enabled systems exhibit three distinctive and emergent characteristics. \textbf{\ding{182}  Ambiguous capability boundaries:} their capabilities are limited by pretrained models rather than explicitly defined functions. Moreover, an understanding of what pretrained models can do is unclear. \textbf{\ding{183} Context sensitive behaviors:} their behaviors heavily depend on prompt design and user input context. Even minor modifications may lead to qualitatively different responses~\cite{errica2024did}. \textbf{\ding{184} Continuous evolution:} Pretrained models are frequently updated through fine-tuning~\cite{hu2022lora}, parameter-efficient adaptation~\cite{hu2022lora} to fix issues, incorporate new knowledge, or improve alignment. Each update may shift system behaviors in unforeseen ways.

These characteristics challenge the foundational assumptions of traditional requirements engineering. Specifically, current mainstream methodologies (\eg waterfall or agile practices~\cite{alshamrani2015comparison}) assume that system behaviors are predictable, specifiable, and verifiable, thereby treating requirements as deterministic specifications~\cite{Jin2023SRE_En}. However, ambiguous capability boundaries hinder stakeholders from defining a precise feature set. The context-sensitive characteristic undermines the assumption that inputs and outputs can be exhaustively enumerated and verified. The evolution characteristic requires continuous monitoring and tracking of model capabilities, prompts, and data sources~\cite{alves2023status}.

We investigate this problem and identify six open challenges that highlight the disconnect between traditional requirements engineering practice and the reality of pretrained-model-enabled systems. \textbf{(1) Breakdown in requirements elicitation:} stakeholders rarely possess a clear and shared picture of a pretrained model's capabilities and limitations, which makes conventional elicitation techniques ineffective~\cite{alves2023status}. \textbf{(2) Intent-capability misalignment:} there is a mismatch between granularity in human intent and model capability description, which makes it difficult to select pretrained models that can satisfy user intent~\cite{pei2022requirements}~\cite{jin2025automatic}. \textbf{(3) Inadequate specification:} natural language requirements specifications lack structural rigor and semantic guarantees, which limits their effectiveness as precise and robust specifications. \textbf{(4) Specification quality assurance gap:} prompts act as requirements specifications, but there is no established method to validate their completeness, consistency or testability, which leaves stakeholders uncertain whether the specifications are fit for purpose~\cite{heyn2025causal}. \textbf{(5) Emergence of under-specified non-functional requirements:} the introduction of pretrained models brings new quality dimensions (\eg hallucination rate, fairness, and explainability) that expand the traditional scope of non-functional requirements, but they still lack common definitions, objective metrics, and mature engineering practices. \textbf{(6) Unmanaged evolution of model artifacts:} prompts, model checkpoints, and contextual configurations change rapidly without robust versioning or traceability, which leads to configuration drift and audit gaps~\cite{jin2025first}. 

\begin{figure*}
    \centering
    \includegraphics[width=0.9\linewidth]{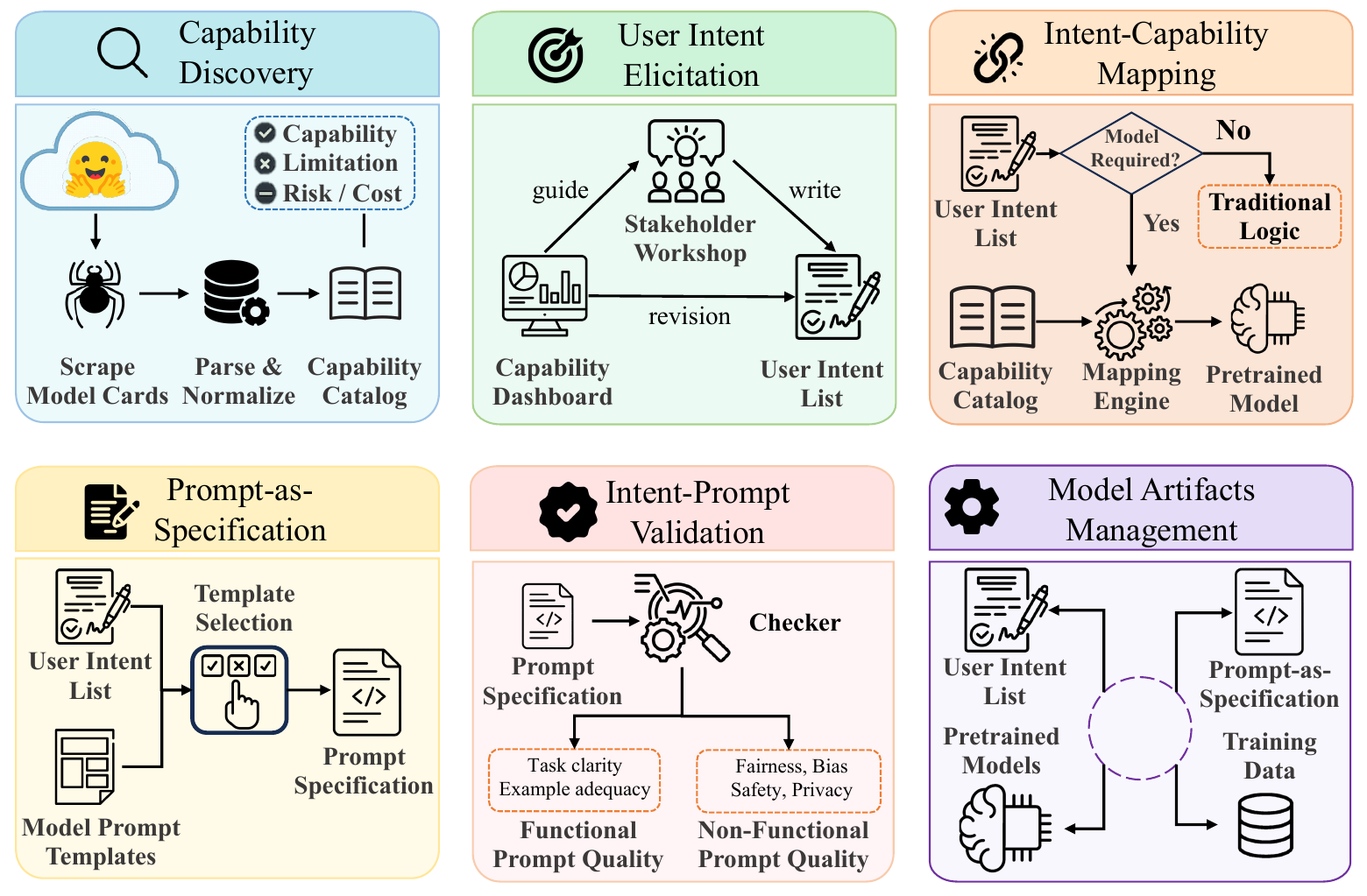}
    \caption{A Conceptual Framework for Requirements Engineering of Pretrained-model-enabled Systems.}
    \label{fig:overview}
\end{figure*}

To address these emerging challenges, we propose a conceptual framework that restructures the requirements engineering life cycle into six interconnected activities, as illustrated in Figure~\ref{fig:overview}.

\begin{itemize}
    \item \textbf{Model Capability Discovery:} uncovers and bounds what pretrained models can do based on their ecosystems (\eg the HuggingFace ecosystem).
    \item \textbf{User Intent Elicitation:} elicits stakeholder goals, constraints, and success criteria through interactive, model capability-driven techniques.
    \item \textbf{Intent-capability alignment:} matches the elicited intents to candidate models whose capabilities best satisfy the desired functionality.
    \item \textbf{Prompt-as-specification:} elevates prompts to first-class and executable requirements specification, giving them explicit structure and semantics.
    \item \textbf{Behavior-driven specification validation:} validates prompt specifications via statistical testing, scenario simulations, and human-in-the-loop reviews to ensure completeness, consistency, and robustness.
    \item \textbf{Model Artifacts Management:} versions and traces prompts, model checkpoints, and fine-tuning datasets to control evolution, prevent configuration drift, and maintain auditability.
\end{itemize}

Building on this conceptual framework, we envision several promising research directions for requirements engineering of pretrained-model-enabled systems. \textbf{(1) Pretrained-model Capability Ontologies:} constructing structured ontologies that formally capture, organize, and evolve pretrained-model capabilities. \textbf{(2) Requirements-Driven Model Selection}: devising methods that align stakeholder intents with candidate models through automated capability matching and recommendation. \textbf{(3) Prompt Specification Languages:} developing syntactically and semantically constrained languages for prompt specifications. \textbf{(4) Emergent Quality Modeling}: defining, measuring, and integrating novel quality attributes (\eg hallucination risk) into requirements engineering artifacts and processes. \textbf{(5) Life cycle Traceability:} establishing robust mechanisms for versioning, provenance, and audit mechanisms for prompts, model checkpoints, and fine-tuning datasets. We hope this work lays the groundwork for a systematic exploration of requirements engineering of pretrained-model-enabled systems.

The remainder of this paper is organized as follows. In section~\ref{sec:challenges}, we discuss existing challenges when traditional requirements engineering methodologies are applied to pretrained-model-enabled systems. In section~\ref{sec:framework}, we present our conceptual framework for requirements engineering of pretrained-model-enabled systems. Section~\ref{sec:directions} outlines promising research directions, and section~\ref{sec:conclusion} concludes the paper.
  
\section{Open Challenges}~\label{sec:challenges}

The characteristics of pretrained-model-enabled systems present challenges not typically encountered in traditional requirements engineering. This section analyzes six concrete challenges, which span the entire requirements engineering life cycle (\ie elicitation, analysis, specification, validation, and management).

\textbf{Challenge 1:  Requirements elicitation under opaque model capabilities.} In traditional systems, stakeholders and engineers collaborate to define requirements based on explicit, decomposable functionalities. However, the pretrained-model-enabled systems' core behaviors stem from latent knowledge encoded in the model rather than predefined features. This opacity makes it nearly impossible for stakeholders to articulate clear boundaries of ``what the system can or cannot do''. Traditional elicitation techniques, such as structured interviews, use-case workshops, or user stories, rely on stakeholders’ understanding of pretrained models' capability. In practice, stakeholders may overestimate model capabilities or underestimate their limitations.


\textbf{Challenge 2: Intent-capability misalignment.} Even when stakeholder intents are partially elicited, mapping them to a suitable pretrained model remains non-trivial. Current model documentation (\eg model cards and technical reports) is often too coarse-grained to align with task-specific stakeholder requirements. For example, a model card might claim  ``strong performance on question answering'' but omit details about its handling of domain-specific jargon for a medical assistant system. This mismatch may result in engineers selecting models that are either overly capable or insufficient for the intended task.


\textbf{Challenge 3: Absence of structured and robust prompt specification template.} In pretrained-model-enabled systems, prompts often serve as ``executable specification'', which directly demonstrates how the model processes inputs and generates outputs. However, unlike traditional software requirements specification documents, prompt specifications lack a standardized structure and a semantic guarantees template. This makes it difficult to ensure that prompt-based specifications are complete, consistent, or traceable to user intent.


\textbf{Challenge 4: Prompt specification quality assurance gap.} Prompts serve as executable requirements specifications, but the community lacks systematic methods to assess their completeness, consistency, and testability. This quality assurance gap leaves stakeholders unable to confidently judge if a prompt is fit for purpose, exposing systems to silent failures, safety risks, and costly late-stage revisions.

\textbf{Challenge 5: Emergent non-functional quality attributes without standard metrics.} Pretrained-model-enabled systems introduce a new class of non-functional requirements that extend far beyond traditional frameworks like ISO/IEC 25010, including hallucination rate and output explainability. These attributes lack objective measurement metrics and guidelines for balancing trade-offs, which may make systems ignore critical stakeholder expectations.


\textbf{Challenge 6: Rapid and multi-factor evolution of model artifacts without governance.} Prompts, checkpoints, and fine-tuning datasets mutate faster than traditional code bases. Each update of these artifacts can silently shift system behaviour, causing configuration drift, and undermining auditability. However, unlike software code, these artifacts lack robust governance. This not only hinders reliability but also violates compliance requirements for accountability.

\section{Conceptual Framework}~\label{sec:framework}

To tackle the challenges outlined above, we propose a conceptual framework that reorganizes the requirements engineering process for pretrained-model-enabled systems into six interrelated activities, as illustrated in Figure~\ref{fig:overview}. These activities are designed to adapt to the unique characteristics of such systems and bridge the gap between traditional requirements engineering practices and the realities of pretrained-model-enabled systems.

\subsection{Model Capability Discovery.}
This step aims to facilitate understanding of what pretrained models are capable of. Specifically, given a pretrained model ecosystem (\eg HuggingFace and Github), we need to systematically identify model capabilities, limitations, risks, and associated costs.

To achieve this goal, this process starts by extracting information from model cards, which typically provide details about a model's intended uses, performance, and known limitations. Then the collected model cards are parsed and normalized to extract relevant capability information. The outcome is a model capability catalog, which is a structured repository that documents the diverse capabilities of pretrained models. This catalog servers as a foundational resource for subsequent activities in the requirements engineering life cycle, enabling stakeholders to gain a clearer understanding of what the models can offer from the start. For example, if a pretrained model is being considered for a customer service chatbot, the capability catalog would outline its proficiency in handling different types of customer inquiries.

\subsection{User Intent Elicitation.}

Once we have constructed the capability catalog to help stakeholders understand model capabilities, the next step is to elicit the user intent of their stakeholders (\eg goals, constraints, and success criteria). This process is conducted using an interactive and capability-driven approach.

Stakeholder workshops play an important role in this process. In these workshops, a capability dashboard, which presents information from the capability catalog in a visual and accessible format, is used to guide discussions. Stakeholders then articulate their intents, which are documented in a user intent list. This list is not static, which undergoes revision as stakeholders gain a better understanding of the model capabilities and how they can be applied to meet business scenarios. For instance, in a project to develop an AI-powered content generation system, stakeholders might initially have a broad intent of ``generating high-quality blog posts''. Through the workshop and interactions with the capability dashboard, this intent may be refined to ``generating blog posts on technology topics within 500 - 1000 words while avoiding plagiarism'', taking into account the model's capabilities in content generation and its limitation in handling highly specialized technical content.

\subsection{Intent-Capability Mapping.}

The next activity is to map the elicited user intents to the capabilities of candidate pretrained models. This involves assessing how well a model's capabilities align with the user intents.

This process begins with the user intent list and the capability catalog. A determination is made on whether a pretrained model is required for fulfilling each user intent item. If no required, the intent can be implemented using traditional rule-based logic. However, a pretrained model is involved in most cases. A mapping engine can be trained to match user intents with relevant capabilities in the catalog. For example, if the user intent is to ``classify images of different animal species with high accuracy'', the mapping engine should match this intent to models in the capability catalog that have demonstrated strong performance in image classification tasks, particularly those related to animal species recognition.


\subsection{Prompt-as-Specification.}

In pretrained-model-enabled systems, prompts play a crucial role as they can be elevated to first-class and executable requirements specifications. This activity involves using the user intent list and model prompt templates to create a structured prompt specification.

Specifically, a library of model prompt templates can be created. To transform high-level user intents to a low-level prompt specification, an appropriate prompt template is chosen based on the intents. Then the selected template is filled with key information to produce a prompt specification, which can have an explicit structure and defined semantics. For example, if the user intent is to ``summarize long legal documents into concise key-point summaries'', a suitable prompt template for legal document summarization would be selected. The resulting prompt specification may detail the format of the input, the expected output, and any specific instructions to ensure alignment with the user intents.



\subsection{Prompt Specification Validation.}

After creating the prompt specification, it must be validated to ensure completeness, consistency, and robustness. This can be achieved through behavior-driven specification validation. 

A checker can be trained using synthesized data or utilize predefined criteria to assess the prompt specification against various quality criteria. This checker aims to evaluate both functional and non-functional quality aspect. Thus, these criteria can include task clarity (\ie whether the prompt clearly defines the task), example adequacy (\ie if there are sufficient and appropriate examples provides in the prompt), as well as functional aspects like fairness, bias, and privacy. Specifically, this may involve statistical testing, scenario simulation, and human-in-the-loop review. For instance, if the prompt specification is for a credit scoring assistant system, the check should verify that the prompt clearly defines the credit scoring task and ensure that the resulting scores are fair across different demographic groups.


\subsection{Model Artifacts Management.}
The final activity focuses on the effective management of model artifacts. These model artifacts include prompts, pretrained models checkpoints, and training datasets.

This involves versioning and traceability across these model artifacts and user intent to prevent configuration drift and ensure auditability. As user intents evolve, changes to the prompts, model checkpoints, and fine-tuning datasets must be tracked. For example, if a prompt is modified to improve the performance of a text-to-image generation system, the new version of the prompt should be recorded, which allows for easy rollback if needed and provides a clear audit trail for compliance and debugging purposes. Additionally, managing the pretrained models and their associated training data ensures that the system remains reproducible and any changes can be traced back to specific artifact modifications. 
\section{Research Directions}~\label{sec:directions}

Building upon the proposed framework, we outline several promising research directions to advance requirements engineering for pretrained-model-enabled systems. Specifically,  we believe that future work can be conducted in-depth from the following aspect. 
\begin{itemize}
    \item Constructing pretrained model capability ontology to systematically capture, organize, and evolve model capability, and explore effective ontology-driven approaches for user intent elicitation and automated alignment with model capabilities.
    \item Designing data-driven methods that map elicited user intents to candidate models. This includes metrics for trade-off analysis and tools to recommend optimal models based on intent-specific criteria.
    \item Creating standardized, syntax-aware prompt languages with semantic guarantees to replace natural language prompts. It can bridge the gap between user intent and executable specifications.
    \item Developing systematic techniques to assess prompt quality, including automated checks for completeness, robustness, and alignment with non-functional requirements.
    \item Designing versioning and traceability frameworks for dynamic artifacts to manage evolution. This includes tools for tracking changes, auditing drift, and ensuring compliance with regulatory requirements.
\end{itemize}

\section{conclusion}~\label{sec:conclusion}
Pretrained-model-enabled systems introduce unique challenges to traditional requirements engineering, stemming from their ambiguous capabilities, context sensitivity, and dynamic evolution. These challenges necessitate a paradigm shift in how requirements are elicited, specified, validated, and managed. Our conceptual framework reorganizes the requirements engineering lifecycle around six interconnected activities—from model capability discovery to artifact governance—tailored to the distinctive properties of pretrained models. By addressing gaps in elicitation, alignment, specification, and evolution, this framework provides a foundation for robust requirements engineering practices in this emerging domain.




\useunder{\uline}{\ul}{}

\normalem
\balance
\bibliographystyle{IEEEtran}
\bibliography{main}

\end{document}